\documentclass[
reprint,
 amsmath,amssymb,
 aps,
prb,
]{revtex4-2}

\usepackage{graphicx}% Include figure files
\usepackage{dcolumn}% Align table columns on decimal point
\usepackage{bm}% bold math
\usepackage{float}
\usepackage[table]{xcolor}
\usepackage{subcaption}
\usepackage{upgreek}
\usepackage{mathrsfs}
\usepackage{braket}

\begin{document}

%\preprint{APS/123-QED}

\title{Density functional description of long-range electron Coulomb interactions in bulk SnS}% Force line breaks with \\

\author{Stefanos Giaremis}
\author{Joseph Kioseoglou}
\author{Eleni Chatzikyriakou}%
 \email{elchatz@auth.gr}
\affiliation{%
 School of Physics, Aristotle University of Thessaloniki, 54124 Thessaloniki, Greece
 %\\
}%

\date{\today}% It is always \today, today,

\begin{abstract}
A high-throughput benchmarking technique for testing the performance of different exchange-correlation functionals and pseudopotentials is proposed and applied to bulk SnS. It is shown that, contrary to the popular view that the local density approximation can best describe layered materials, a semilocal pseudopotential with a functional having a gradient dependence better described lattice vectors and `tetragonicity' of the lattice. We classify the pseudopotentials based on this value and show that the participation ratio of maximally localized Wannier functions follows the theory which states that more distorted structures have higher anti-bonding hybridization as stabilizing factor. In order to classify pseudopotentials, the local and nonlocal potential contributions to the dynamical Born effective charges are taken for each pseudopotential. Finally, a strategy is proposed for learning exchange-correlation functionals based on the distinction between short and long range parts of the Kohn-Sham potential.
%\begin{description}
%\end{description}
\end{abstract}

\maketitle

%\tableofcontents

\section{Introduction}

	Significant advances in first principles studies have recently opened new opportunities for the examination of quantum phenomena in physics and chemistry at a very detailed level and reduced cost \cite{Hermann2022,Carleo2019}. One important issue that arises is the assessment of each technique, in terms of various factors such as speed and accuracy \cite{Wu2023}.
	
	High-throughput methods is another category that has gained popularity \cite{Vitale2020, Gresch2018, Gresch2017}. Beside analytical solutions used to screen large sets of proposed compounds \cite{Olsen2018}, other works use the same pseudopotentials, or pseudopotential families, applied to all examined compounds, in order to find the desired solutions. Another alternative is to use high-throughput in conjunction with machine learning techniques, with different applications including the detection of ferroelectric materials and testing of appropriate descriptors \cite{He2021}. Here we follow a different approach, in which we use high-throughput techniques in order to benchmark the pseudopotentials themselves. We do this following a physically-motivated procedure that lays the grounds for machine learning dynamical lattice properties, without the use of separate classical molecular dynamics run, but in combination with first principles electronic properties in Density Functional Theory (DFT). This combined calculation is also more convenient in the case of layered compounds, where the long-range interactions play a significant role both for both structure and properties, such as in twisted multi-layers.
	
    Due to limitations of the exchange-correlation (XC) functionals in realistically simulating interactions of different range orders \cite{Polo2003}, internal properties such as binding energies, polarization and phonon effects can present a qualitative difference in the amount of error introduced compared to more `combined' quantities, such as, lattice constants. In SnS, the internal pressure and forces at different directions that result from the involvement of the lone pairs, makes it hard to give accurate quantitative results. For example, the free-standing 2D systems, can be close to a phase transition and therefore predicted to be stabilized with the addition of a substrate, but it is uncertain whether a different pseudopotential (PP) would produce similar results \cite{Malyi2019}. In this work, we benchmark pseudopotentials for their ability to describe this process in SnS, however, our results can be generalized to other types of compounds containing lone pairs, such as post-transitions metal (Tl, Pb, B), or other types of chalcogenide-based compounds (i.e., GeSe, SnSe).
    
    \subsection{Crystal symmetry}
    \label{sec:symmetry}
    
    SnS has many intriguing properties in its 2D form, however, the bulk form has also shown important properties that include room-temperature valley physics \cite{Lin2018, Chen2018, Chatzikyriakou2019}. In this work, we examine the bulk form in order to avoid the depolarization field effects that arise from the periodic nature of Density Functional Theory. Although there is enough degree of general consensus over its crystal structure, the details have been controversial until very recently \cite{Burton2012, Biacchi2013}.
    
\begin{figure}
   \centering
     \includegraphics[width=.40\textwidth]{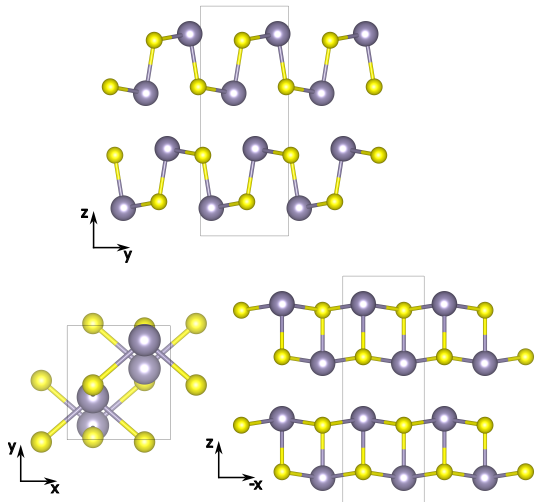}
     \caption{Crystallographic direction notation used in the text.}
         \label{fig:structure}
   \end{figure}

    Sn(II) compounds like SnS form covalent bonds with mixed ionic character \cite{Rundle1964}. In this text, we follow the notation of the  direction being perpendicular to the separate layers, and the x and y being the in-plane directions (fig. \ref{fig:structure}). In this configuration, the bonds are formed between Sn and S orbitals of the same layer in both the in-plane and out-of-plane directions. Sn lone pairs extend in the void between, what appears to be, weakly bonded layers.
    
\begin{figure}
   \centering
     \includegraphics[width=.46\textwidth]{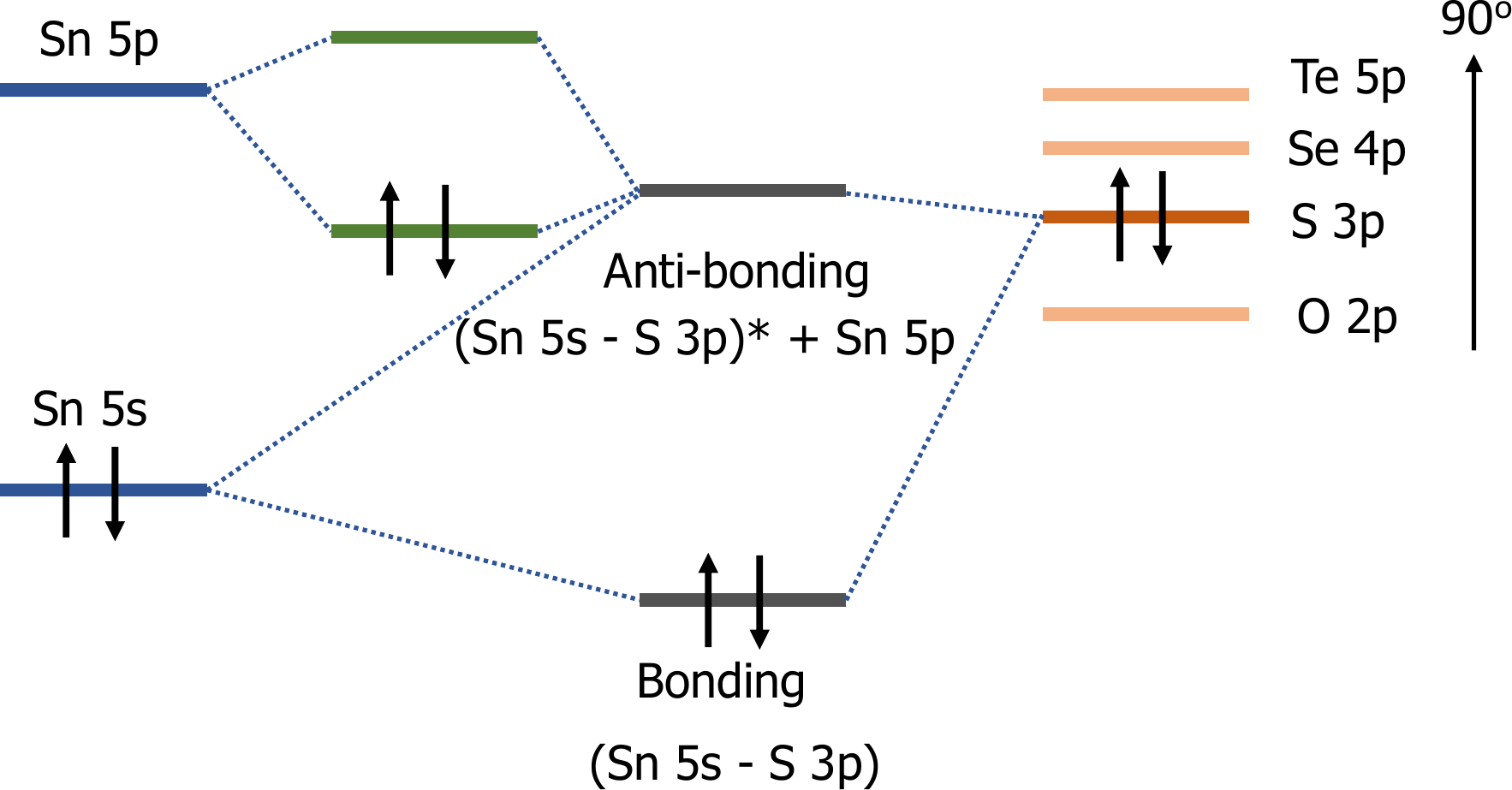}
     \caption{Energy level diagram for Sn(II) bonding according to the RLP model \cite{PrInorgChem8,Walsh2011}. The Sn 5p states hybridize with the (Sn 5s - anion p) anti-bonding states.}
         \label{fig:LRP_energy}
\end{figure}

    It was realized from early on that lone pair formation was important for the lattice structure. In the revised lone pair (RLP) model \cite{PrInorgChem8,Walsh2011}, this view persists\cite{Payne2006}. Distortions of the lattice are due to the hybridization of the Sn(5p) with the Sn(5s)+S(3p) anti-bonding orbitals (Figure \ref{fig:LRP_energy}). It is generally assumed that the more the energies of the latter overlap, the more strongly they interact, producing stronger anti-bonding states, which hybridize with Sn(5p) states. This creates a higher distortion, and a larger interlayer distance. The resulting lattice is less symmetrical, with angles away from 90\textsuperscript{o} (e.g. SnO). Weaker interactions (SnS, SnSe), and eventually breaking of such bonds (SnTe), leads to more symmetrical structures \cite{Walsh2005}. 
    
    Similar symmetry-breaking transitions also occur as a function of temperature of bulk SnS from the higher-temperature, higher-symmetry Cmcm spacegroup to the the low-temperature, lower-symmetry Pnma \cite{VonSchnering1981, Lebedev2018}, and the same orbital hybridization model is believed to be behind this mechanism \cite{Li2015}.
    
    While these facts are well-known, and many authors choose to not include vdW effects in their calculations \cite{Dewandre2019}, what is generally not mentioned often is the important role the stereochemically active lone pairs play at stabilizing the structure. These anti-bonding orbitals, produce a stabilizing force that is stronger than the weak vdW forces \cite{Song2018, Xu2022}, and they are also the reason single-crystal 2D SnS is hard to achieve experimentally on a large scale \cite{Ehm2004,Higashitarumizu2020,Tian2017,Li2017,Xia2016,Sarkar2022}. 
    
   It is also worth mentioning that recent studies showed that the RLP model is not fully applicable to the case of SnTe, as well as other cases such as (PbSe, PbTe, GeTe). In these cases, the type of bonding has been named metavalent, and is intermediate between metallic and covalent or ionic bonding \cite{Ronneberger2020}. While, in principle there materials should also have stronger interlayer bonds than vdW type, the atoms share approximately one electron between them \cite{Raty2021}. The region where the electron extends and its localization properties depend on many intrinsic (e.g. dimensionality) as well as extrinsic factors.
     
    \section{Density functional approximation}     
    
    The starting point of our analysis is the Kohn–Sham potential, whose general expression is,
      
      \begin{equation}
      V_{KS} = V_{ion} + V_{H} + V_{XC}
      \label{eq:Vk_s}
      \end{equation}
      
      where the terms correspond to the electron-ion interaction, Hartree and XC potential energies respectively. This, along with the Hohenberg-Kohn theorem, relates the energy with the density of the system and allows for the separation of the contributions of each type of potential. The kinetic energy of the electrons in inferred from the charge density that is produced from this potential energy after the self-consistent procedure.
     
    \subsection{Exchange-Correlation functional}
     
     Starting with V\textsubscript{XC}, we note the important difference between the performance of the Generalized Gradient Approximation (GGA) functional and the Local Density Approximation (LDA). Short-range electron correlation effects are explicitly included in the DFT functionals, however, long-range correlations are also thought to be included through the self-interaction error. The underestimation of the approximated exchange hole (that depends on the exchange functional) is equivalent to a re-construction of the bands around the reference electron and mimicking nondynamic, long-range correlation effects \cite{Polo2003}. 
     
     In general, the LDA exchange hole is diffuse and overestimates the correlation energy. It produces larger self-interaction errors than GGA for the core and lone pair orbitals. At the same time, the GGA hole is more localized than the HF exchange hole in the bond and lone pair regions due to the Coulomb self-repulsion being over-compensated \cite{Polo2003}. 
     
     It is generally established that the majority of the GGA functionals exaggerate bond length, while LDA `overbinds'. LDA functionals are thought to better describe layered materials, that is, a description that involves regions of low electron density \cite{Marini2006}, and this has been adopted for SnS as well \cite{Lebedev2018, Tritsaris2013, Vidal2012}. There is a second factor that affects accuracy for materials with polarization properties, that is the insensitivity of the local functional to the polarization, when the exact XC functional is polarization dependent \cite{Polo2002, Gonze1995, Ghosez1997}. This, however, becomes a problem only when the electronic response becomes equally significant as the ionic response, which is many times not the case in ferroelectric oxides.
     
     As described in the previous section, stabilization of SnS is a combination of the hybridization between the bonding and the anti-bonding orbitals that reside at distances where the approximation methods produce qualitatively different results. This means that there exist contributions from both bonding and non-bonding regions, while the long-range vdW interaction terms play a secondary role. At the same time, it pays to consider the lattice and electronic contributions to the dielectric response of the system, and the quantities produced by each approximation method.
     
     In the self-interaction-corrected  (SIC) approximation, the exchange-correlation energy of a single, fully occupied orbital is taken to cancel exactly the self-Coulomb energy (the functional of the electron number density) \cite{Perdew1981}. While this has been shown to improve some total and ionization energies in some systems, it has also been shown to worsen the results in a plethora of systems that need the self-interaction error in order to compensate for the lack of long-range force description in DFT \cite{Polo2003}. In all cases, the SIC effects are reduced as a function of distance from the ion \cite{Perdew1981}. In our case, for the 5s and 5p orbitals of Sn this effect might become negligible.
     
     In order to keep the computational time and general simulation complexity low, only approximations that belong to the first rungs of the Jacob's ladder were used. For LDA, Slater-type exchange \cite{Slater1951} and Perdew–Wang (PW) correlation energy \cite{Perdew1992} are assumed, while for GGA, the Perdew, Burke, Ernzerhof (PBE) gradient correction to the exchange and correlation \cite{Perdew1996}. The solids-oriented PBEsol, with the diminished gradient dependence \cite{Perdew2008} was also considered, as well as the SIC Perdew Zunger (PZ) LDA functional \cite{Perdew1981}.
     
     \subsection{Electron-ion potential}
     
     Four basic types of approximations of the electron-ion interaction potential energy, V\textsubscript{ion}, have been examined. Starting with the semilocal (SL) PP,
     
     \begin{equation}
     \begin{split}
     	V_{ion} = V_L + \sum_l V_{SL,l} \hat{P}_l \\
     	V_{SL,l} = V_{l,NL} - V_{L}
     \end{split}
	\end{equation}
     
     where the subscript L denotes local pseupotential, NL the non-local potential and l the angular momentum component of the wave-function. For a nonlocal ionic potential, a different pseudopotential, V\textsubscript{l}, should be acting on each angular momentum component,
     
     \begin{equation}
     V_{NL} = \sum_{lm} \bra{\Psi_{lm}} V_{l} \ket{\Psi_{lm}} 
     \end{equation}
     
     where $\Psi_{lm}$ are the spherical harmonics. Semilocal PPs essentially become non-local only beyond a cutoff radius, a process which allowed the plane-wave basis set to be expanded to the nonlocal description by reducing the computational cost, but is also a source of error.  
        
     SL PPs are norm-conserving, that is, the norm of the pseudo wavefunction is conserved inside the core region \cite{Troullier1991}. Another form of the norm-conserving PPs was given by Rappe, Rabe, Kaxiras, and Joannopoulos (RRKJ) \cite{Rappe1990, Rappe1991} who expanded each pseudo-wave-function as a linear combination of basis functions. The implementation of this PP in the present work is ultrasoft (US). 

     For PAW PPs, we use the KJ derivations \cite{Kresse1999}. In the PAW formalism, the atomic-based projectors $\beta$ participate in the KS potential during the self-consistent cycle through $\sum_i\braket{\psi_i|\beta_n}\braket{\beta_m|\psi_i}$, with $\psi_i$ the calculated wavefuction \cite{Giannozzi2009}.

     In Ref. \cite{Hamann2013}, Hamman also proposed the optimized version of the norm-conserving Vanderbilt PPs, which rely on the multiple projector scheme proposed by Vanderbilt for passing from norm-conserving to ultrasoft PPs. These are the Optimized Norm Conserving (ONCV) used here. The details of the implementations are not important for this work, because their differences can be captured in the local and non-local definition of the Born Effective charges given in the next section.     
      
     \subsection{Range of interactions} 
     \label{sec:range}
   
The existence of ferroelectricity in displacive ferroelectrics, results from the interplay between short-range and long-range interactions \cite{Wang2012, Cochran1960}. That is, only short range interactions are required to support the high-symmetry paraelectric phase, whereas long-range interactions are required, in order for the ferroelectric phase to persist in any configuration (when raising the temperature, thinning down to a few layers or doping). If the long-range interactions are somehow screened, ferroelectricity can, in principle, be suppressed, even above the transition temperature \cite{Wang2012}. Here we examine the effects of pseudopotential in the description of lattice dynamical properties, as they arise from differences in the short and long range interactions of the electrons and ions of SnS.  

The ferroelectricity results from the broken centrosymmetry in odd number of 2D SnS. In the literature, 2D SnS is considered displacive \cite{Fei2016}, while the defining factor of the ferroelectric/paraelectric properties is the short-range interaction between the lone pairs of electrons \cite{Lebedev2018}. Here we review the quantities involved in the calculations in a first principles set-up. 

Taking each term in equation \ref{eq:Vk_s}: V\textsubscript{XC} is short ranged, V\textsubscript{H} has a short and a long-range part and V\textsubscript{ion} is long (infinite) range. The later is further divided in local and non-local part, V\textsubscript{L} and V\textsubscript{NL} respectively. Of the two, the non-local potential is short ranged, while the local potential has both short-range (SR) and long-range(LR) contributions \cite{Otani2006, Sohier2017}. Therefore,

       \begin{equation}
     \begin{split}
     	V_{KS}^{SR} = V_L^{SR} + V_H^{SR} + V_{XC} + V_{NL} \\
     	V_{KS}^{LR} = V_L^{LR} + V_H^{LR}
     \end{split}
     \label{eq:SR-LR}
	\end{equation}
	
	In DFT codes, the convention is that the SR and LR parts are dictated by the cut-off radius during construction of the PP, and then participate in the Ewald summation with the former being calculated in real space, while the later in reciprocal space. 

     \subsection{Dynamical properties} 
     \label{sec:dynamical}
   
	Independent of the method used to derive them, the Born effective charges, 
	
	\begin{equation}
	Z^* = - \partial \mathbf{F} / \partial\mathscr{E}
	\label{eq:Z}
	\end{equation}
	
	can be divided so as to only take into account contributions of a specific potential. The forces are related to V\textsubscript{ion},
	
	\begin{equation}
	F_{\alpha,i} \propto \int_{\Omega}n\left(\mathbf{r}\right)\frac{\partial V\textsubscript{ion}}{\partial \mathbf{u}_{\alpha,i}}
	\label{eq:forces}	
	\end{equation}
	
	for an atom $\alpha$ in the direction i and $\Omega$ the volume of the unit cell. It should be noted that there is also an ion-ion contribution, as well as a kinetic energy contribution in the calculation of the energy of the system. The kinetic energy is short-ranged, but it is not included as a separate term in the Kohn-Sham potential, but rather deduced from other quantities during the self-consistent procedure. The ion-ion contribution only depends on the atomic numbers of the elements and therefore should be similar for all PPs. 

    \section{Relaxing forces}
    \label{sec:relaxation}
    
   We performed the calculation in the Pnma phase of bulk SnS, with 4 Sn and 4 S atoms. The zigzag and armchair directions are along the \textit{a} and \textit{b} axis, respectively. First we calculated the relaxed bulk structures using each PP. The results are presented in Table \ref{tab:table1}. They are listed in increasing value of the \textit{a/b} ratio. The value that most closely matches the experiment was given by the SL PBE. The rest of the approximations gave values either lower or higher, but with a significant difference from the PBE SL. 
    
    The LDA PPs produced higher ratios than GGA. This can be explained immediately by the property of LDA to overbind. In fact, the more tetragonal structures also have smaller in-plane lattice vectors, with the armchair direction showing more pronounced reduction. The values given by the reduced gradient dependence PBEsol, lie between those of LDA and PBE. Interestingly, the self-interaction corrected LDA (PZ) produced identical \textit{a/b} ratios with pure LDA (PW), however, there was one significant difference: PW reduced the inter-atomic distances of non-bonded Sn-S pairs, reaching bonding lengths, which effectively changed the coordination of the atoms to five-fold. The only other difference is an accompanying slight reduction in the $c$ lattice constant produced with the PW. Overall, the SIC, had the effect of slightly elongating the interatomic distances of the structure, compared to the nonphysical result produced by the LDA.
    
     The armchair (\textit{b}) direction has the largest deviation between pseudopotentials, which can be intuitively understood from the absence of bonds in this direction that results in lower electron density, in which DFT has been notoriously bad at approximating.
    
    The scenery changes slightly when we consider the out-of-plane lattice vectors. Here, the SL PP performed only marginally better than LDA, with the PBEsol giving the best estimate. Again the distances between atoms (interlayer this time) produced by LDA, are smaller than those of PBE. This result shows that, contrary to popular view, the interlayer distances given by LDA are not necessarily better than those of GGA. The PBE results of SL and PAW for the \textit{c} lattice vectors improved with the addition of vdW interaction, however, they remained qualitatively similar.

\begin{table*}%The best place to locate the table environment is directly after its first reference in text
\begin{center}
\caption{\label{tab:table1}%
Primitive lattice constants along the \textit{a}, \textit{b} and \textit{c} directions after relaxation of bulk SnS without vdW forces included, compared to experimental measurements. The value that most closely matched the experiment is given in blue. The results are arranged in increasing \textit{a/b} ratio.
}
\begin{ruledtabular}
\begin{tabular}{ccccc}
\textrm{}&
\textrm{\textit{a} [\r{A}] (\%)}&
\textrm{\textit{b} [\r{A}] (\%)}&
%\multicolumn{1}{c}{\textrm{Decimal}}&
\textrm{\textit{c} [\r{A}] (\%)}&
\textrm{\textit{a}/\textit{b} (\%)}\\
\colrule
exp. \cite{Ehm2004} & 3.987 & 4.334 & 11.199 & 0.919 \\
\colrule

PBE PAW & 	4.027(+1.0) &  4.454(+2.7) & 	11.464(+2.3) & 0.904(-1.7) \\
PBE ONCV & 	4.029(+1.0) &  4.450(+2.6) & 	11.426(+2.0) & 0.905(-1.5) \\
PBE US & 	4.028(+1.0) &  4.440(+2.4)	& 	11.441(+2.1) & 0.907(-1.3) \\
\textcolor{blue}{PBE SL} & \textcolor{blue}{4.023(+0.91)} & \textcolor{blue}{4.360(+0.5)} & \textcolor{blue}{11.362(+1.4)} & \textcolor{blue}{0.922(+0.3)} \\
PBEsol US & 3.989(+0.05) &	4.238(-2.1) &	11.107(-0.8) & 0.941(+2.2) \\  
PBEsol PAW & 3.989(+0.07) & 4.234(-2.2) &	11.106(-0.8) & 0.942(+2.4) \\
PZ US & 	3.961(-0.6) &  	4.189(-3.3) &	10.995(-1.8) & 0.945(+2.7) \\
PZ PAW & 	3.961(-0.6) &  	4.187(-3.3) & 	10.999(-1.7) & 0.945(+2.7) \\
PW ONCV & 	3.957(-0.7) &  	4.182(-3.4) & 	10.975(-1.9) & 0.946(+2.8) \\
\end{tabular}
\end{ruledtabular}
\end{center}
\end{table*}

 The case of under-convergence of wavefunction kinetic energy cut-off, $e_\mathrm{T}$, in terms of the resulting internal pressure was also examined. For some ferroelectrics, transitions have been reported at -5 GPa \cite{Vanderbilt2003}. For SnSe, it has been proposed that the Curie temperature can be tuned up to a few hundred Kelvin, only by the application of a small strain, on the order of 1\% deviation of the lattice constant in the armchair direction \cite{Fei2016}. The differences in different pseudopotentials in describing internal properties such as pressure have been widely examined in the literature for different types of ferroelectrics. For example, in PbTiO\textsubscript{3}, a supertetragonal structure was obtained either with GGA PPs, or with LDA at negative pressure \cite{Vanderbilt20032}.
 
    \begin{figure*}
   \centering
     \includegraphics[width=.69\textwidth]{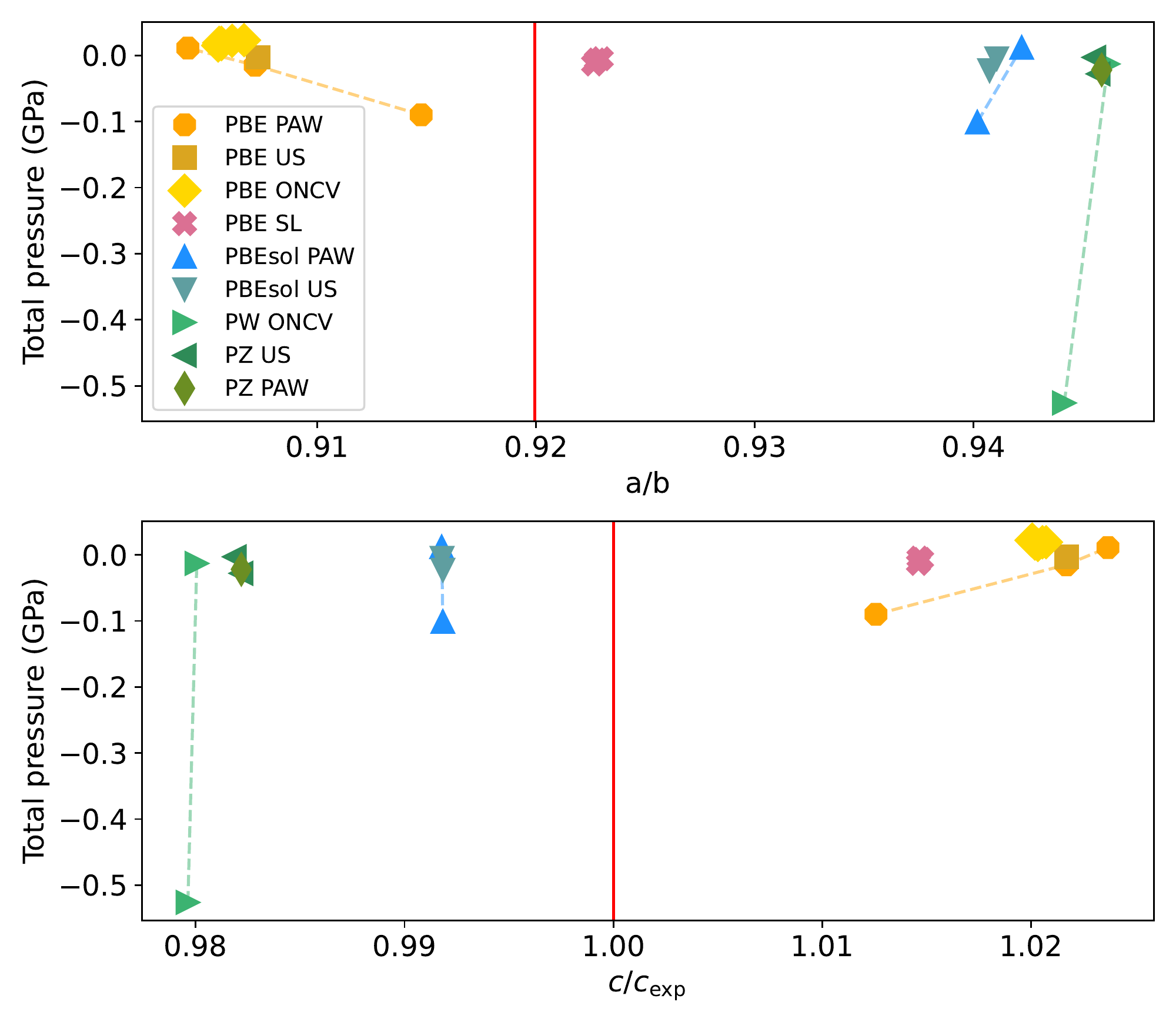}
     \caption{Internal pressure vs. \textit{a/b} lattice constant ratio (top) and \textit{c} lattice constant (bottom). Different points represent different kinetic energy cutoff, $e_\mathrm{T}$ for the wavefunctions. The energy cut-off for charge density, $e_\mathrm{\uprho}$, is 4 $\cdot\ e_\mathrm{T}$ in all cases, except for ultrasoft pseudopotentials, where $e_\mathrm{\uprho} = 8\cdot e_\mathrm{T}$. The solid vertical line shows the experimental value. Unless stated as `PAW', the plane-wave method is used. `PW' refers to the LDA PW functional \cite{Perdew1992}.}
         \label{fig:pressurePP}
\end{figure*}

 The results of the relaxation are shown in fig. \ref{fig:pressurePP}. Some approximations (i.e. PBEsol) generally require higher values of $e_\mathrm{T}$. Lower values produced higher negative internal pressures, but most converged to positive values close to zero at higher $e_\mathrm{T}$. The values of \textit{a/b} ratios, as well as \textit{c} lattice constants only slightly changed for the different values of pressure, but the changes were significant for PBE PAW, which required $e_\mathrm{T}\ >$ 75 Ry to achieve convergence. Some authors tend to use the experimental lattice constants for their self-consistent calculations. In our simulations, there were noticeable internal pressures when the lattice constants used for the self-consistent cycles did not equal to those of the relaxed structure, so we did not adopt this technique anywhere in this work.
 
 Figure \ref{fig:pressurePP} (bottom), further clearly shows that lower \textit{a/b} ratios produce higher interlayer distance, as is intuitively expected for more distorted structures.
    
    \section{Effects of bonding on structure}
    
    \subsection{Projected Density of States}
    
\begin{figure}
   \centering
     \includegraphics[width=.47\textwidth]{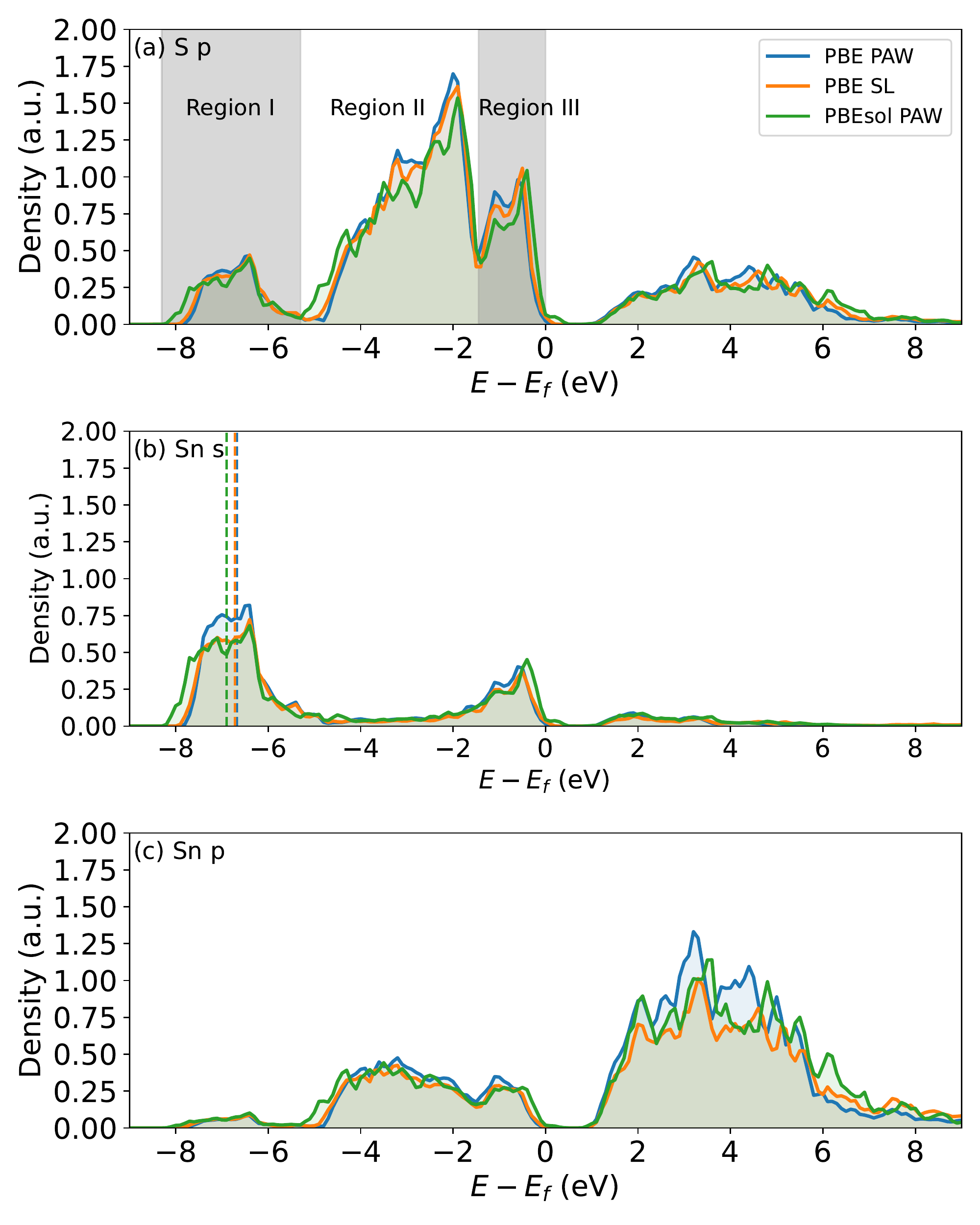}
     \caption{Orbital-projected density of states for (a) the p orbitals of S and (b) the s orbital of Sn and (c) the p orbitals of Sn, using three representative pseudopotentials. The dashed lines in (b) show the centroid of region I for each case.}
         \label{fig:PDOS}
\end{figure}

    To analyze the effects of bonding on the final structure produced, we used Density of States calculations, projected on selected orbitals and sums thereof (PDOS). The results are shown in fig. \ref{fig:PDOS}. This type of representation does not distinguish between bonding and anti-bonding states, however, it reveals details on the overlap of certain orbitals in their energy range. We present only the results of three representative PPs. For the rest, the results are almost overlapping to each case presented here, and their similarities follow the trend presented in fig. \ref{fig:pressurePP}. They are all included in the supplementary material.
    
    We follow the notation of Ref. \cite{Walsh2005} and divide the valence energy range in regions I, II and III (fig. \ref{fig:PDOS}). As it is also visualized in fig. \ref{fig:PDOS}(a), region I is the bonding region, region III is the anti-bonding region. In region II, there exist almost none of the Sn(s) states. The energy difference between them and the Sn(p) states, does not allow the two to directly hybridize. They interact through mediation from the S(p) states \cite{Walsh2005}.
    
    Theory states that stronger anti-bonding hybridization produces larger distortions, and lower \textit{a/b} ratios. We do not present any quantitative results on the integrals here, as it is not easy to set clear boundaries between the regions. We leave this for the next section, where we present hybridization based on the Wannier results. Here, we focus first on the energy ranges. 
    
    In region I, the S(p) states have almost matching energy overlap for each PP used, except for a slight spread-out in energy with PBEsol. At the same time, the S(p) states for PBE PAW have higher density in region III, getting progressively lower for PBE SL and PBEsol. This shows that there are less S(p) states in the anti-bonding region as we go towards more symmetrical structures. 

    By looking at the centroid of the Sn(s) states in Region I in comparison with the peak at the top of region II S(p) states, we can observe the energy distance between them, which points to the amount of hybridization \cite{Walsh2005}. The energy difference is roughly constant, and increases slightly for the more symmetrical structure produced by PBEsol PAW, revealing a less hybridized character of the states. Overall, the density of S(p) in region II gets lower as we go from more distorted (PBE PAW) to more symmetrical structures (PBEsol), the same an in region III.
    
    Contrary to what was observed for the S(p) states in regions III and II, the Sn(s) and Sn(p) states in all regions are almost identical for the three PPs, except for one significant difference: With PBE PAW, the Sn(s) states have a higher density than with PBE SL in region I, while for PBEsol they spread to lower energies. 
    
    To sum up, we observe that while there are more S(p) states in the anti-bonding energy region with increasing distortion, Sn does not seem to follow the same trend. There are, however, differences in the Sn(s) states in the bonding region. 
    
\subsection{Wannier functions}

We convert the DFT states into maximally localized Wannier functions (MLWF), starting from projected WF on the valence s and p orbitals of the Sn and S atoms, resulting to overall 32 bands. For the case of  PBE SL, all valence states were included, for the rest of the PPs, the Sn 4d states were excluded from the wannierization using a disentanglement window. The details of the convergence and checking of the WF results are given in the methods section.

Another method to examine the state would be the Density Overlap Regions Indicator \cite{DeSilva2014}, however, this did not produce valuable results, possibly because it is mostly oriented towards non-bonding orbitals. A third method exists, that is also suitable to high-throughput calculations, as it is devoid of the need to specify an energy window and other parameters. It is based on the `selected columns of the density matrix' (SCDM) algorithm \cite{Damle2014,Vitale2020} and is able to derive Wannier functions at the highest electron density k-points. Although such a method would allow for significant increase in automation, the WFs produced depend on the localization properties of the states, that differ for each PP, as shown below.

\begin{figure}
   \centering
     \begin{subfigure}[b]{0.43\textwidth}
         \centering
         \includegraphics[width=\textwidth]{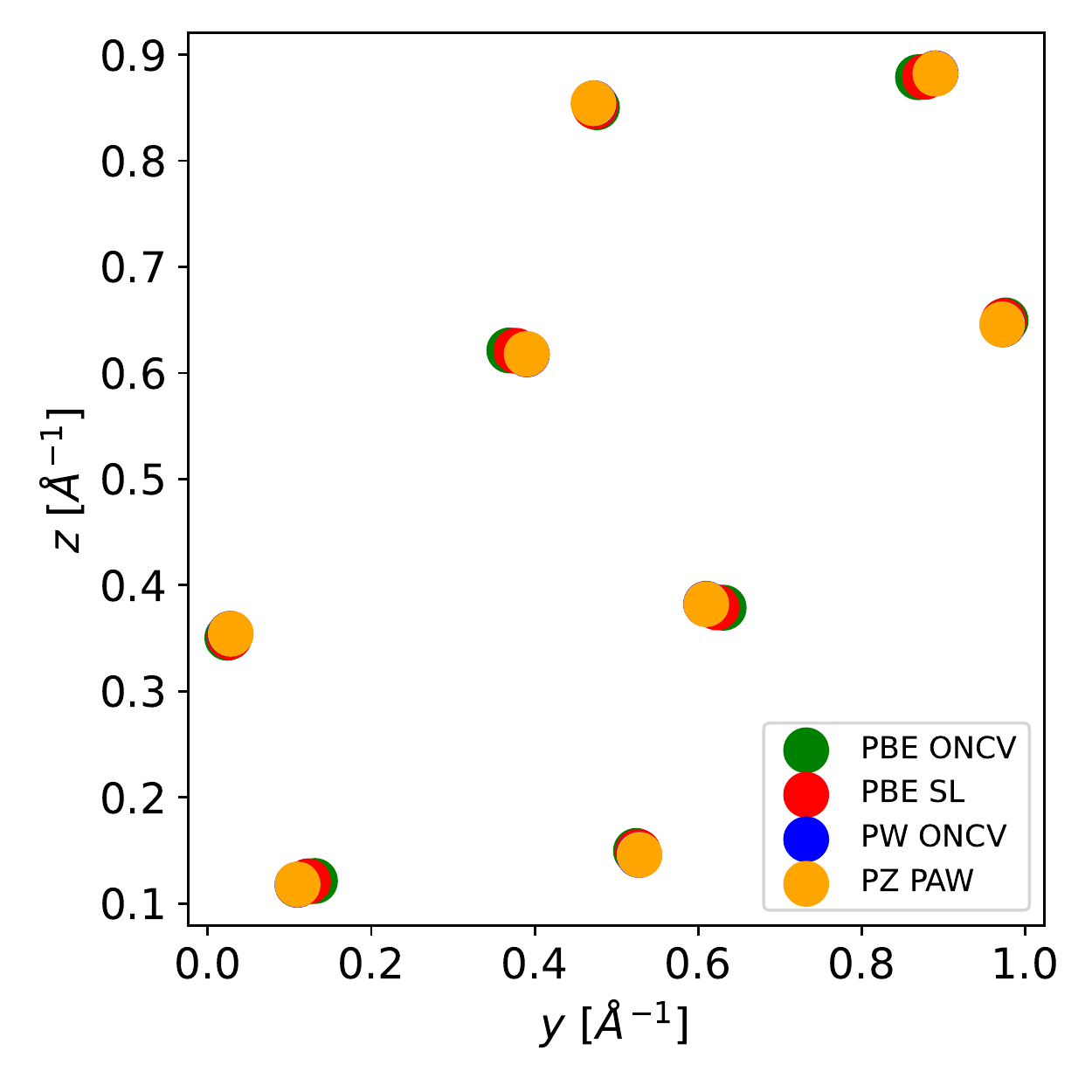}
         \caption{}
         \label{fig:ion_pos}
     \end{subfigure}
    \centering
    \hfill
     \begin{subfigure}[b]{0.49\textwidth}
         \centering
         \includegraphics[width=\textwidth]{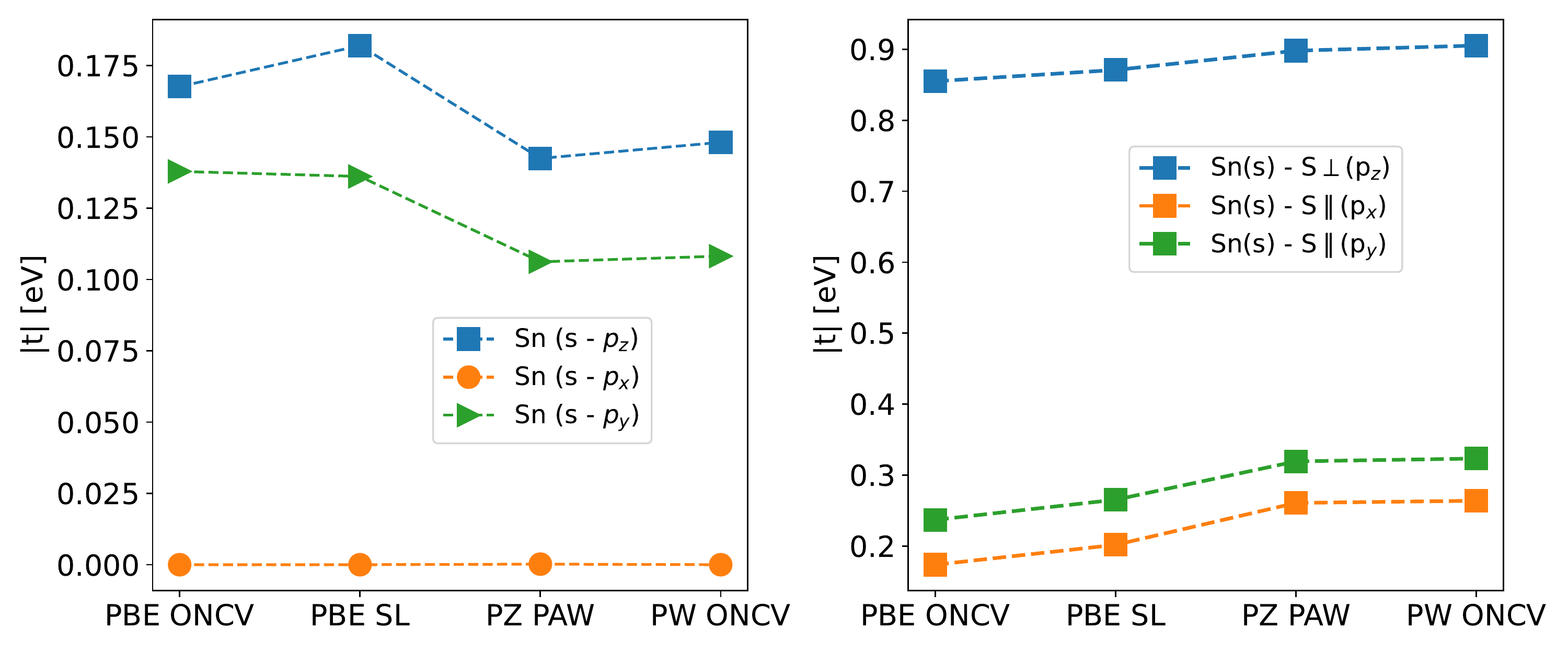}
         \caption{}
         \label{fig:tpair_detail}
     \end{subfigure}
     \caption{(a) Positions of the Sn and S atoms in the unit cell (b) Absolute value of the hopping energy of intra-atomic bonding of Sn (left) and inter-atomic bonding with the neighboring S atoms in the in-plane and out-of-plane directions in the home unit cell (right). The integrals are taken from projected WFs, before the minimization procedure. All Sn-S pairs shown are bonded, except for Sn-S($\parallel$), which is a non-bonding pair.}
\end{figure}

In detail, the problem is related to the `projectability' of the states. Orbitals with more nodes, are harder to separate from the higher and lower bands \cite{Vitale2020}. Therefore the lack of an energy window specification for this method becomes a disadvantage. Indeed, we encountered both problems when trying this method on SnS. To overcome this issue, either the parameters of the SCDM, or the number of bands can be changed. We believe that the second option would be more efficient as only a few conduction bands are required for hybridization with the anti-bonding states, and the states become more `projectable', the lower we are in energy (see supplementary information).

In order to automate the disentanglement procedure, the number of states inside the energy window of interest was calculated from the first principles bands calculation at each \textbf{k} point. The energy range was then re-adjusted so that it becomes equal to the target WFs \cite{Gresch2017, Gresch2018}. Multiple checks were performed regarding the validity and accuracy of the WFs. Without using the minimization procedure, the bonding between the WFs, as revealed by the hopping integrals, follows the trend in the bonding distance. In fig. \ref{fig:ion_pos} the position of the atoms in the $y$ vs. $z$ directions after relaxation using DFT is plotted. The Sn atoms are more displaced towards the direction of the lone pairs as we move to more symmetric structures. The bond lengths expand and contract accordingly.

In fig. \ref{fig:tpair_detail} the hopping integrals between selected orbitals are plotted. The atoms in the top and bottom layers are related to each other through inversion symmetry. Similarly, for the atoms in each layer separately. The bonding integrals are related to each other through such symmetries and the integrals repeat accordingly, with the addition of a ($\pm$1 factor). Consequently, we only show hoppings between orbitals of the top layer Sn and S atoms. It is observed that as the symmetry of the structure increases, intra-atomic hopping in the Sn atom decreases, in favor of inter-atomic hopping energy, which increases. This reflects decrease in distances between the Sn and S atoms with increasing symmetry.

\begin{figure}
   \centering
     \begin{subfigure}[b]{0.49\textwidth}
         \centering
         \includegraphics[width=\textwidth]{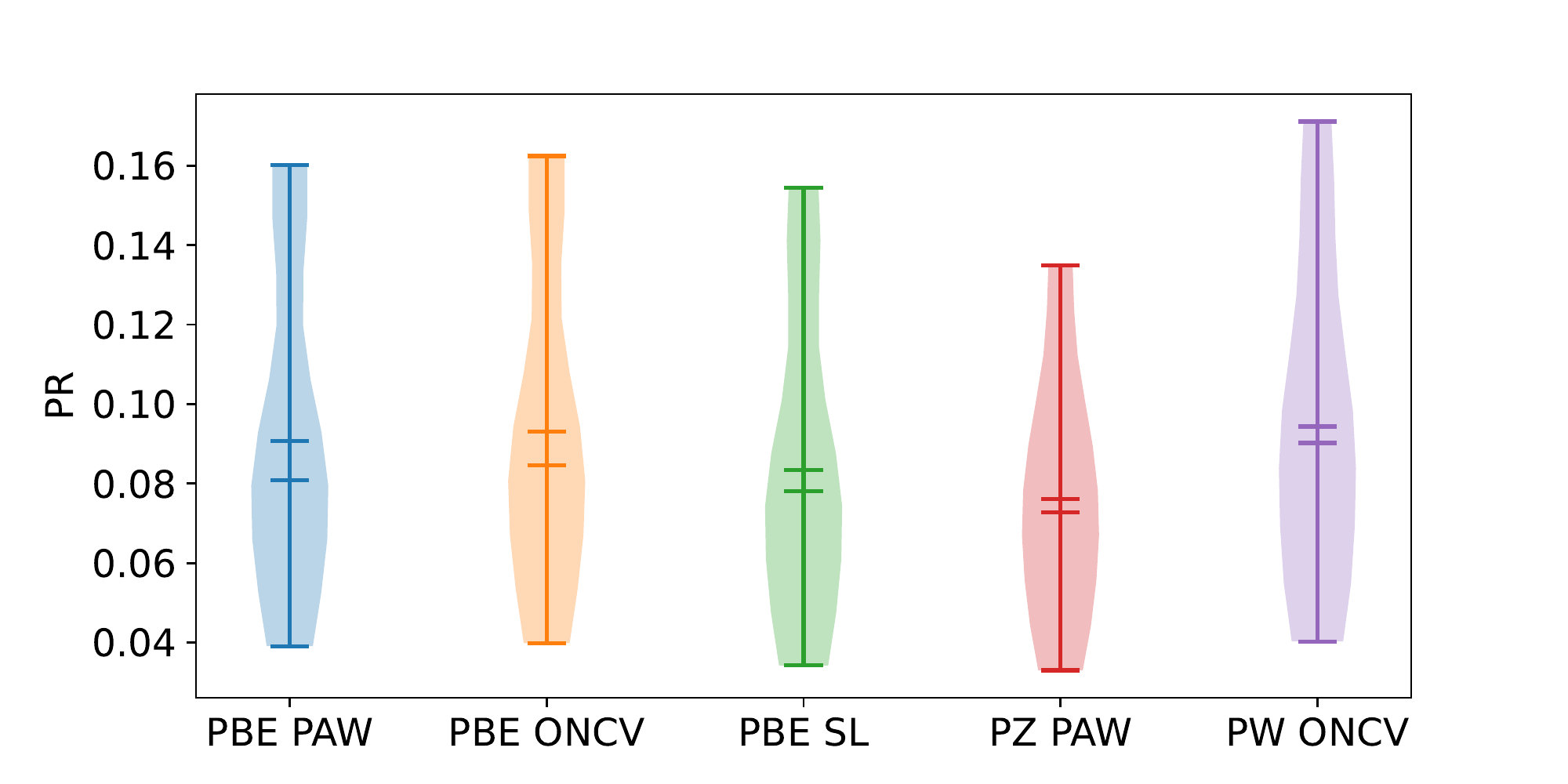}
         %\caption{}
         \label{fig:PR_all}
     \end{subfigure}
    \centering
    \hfill
     \begin{subfigure}[b]{0.49\textwidth}
         \centering
         \includegraphics[width=\textwidth]{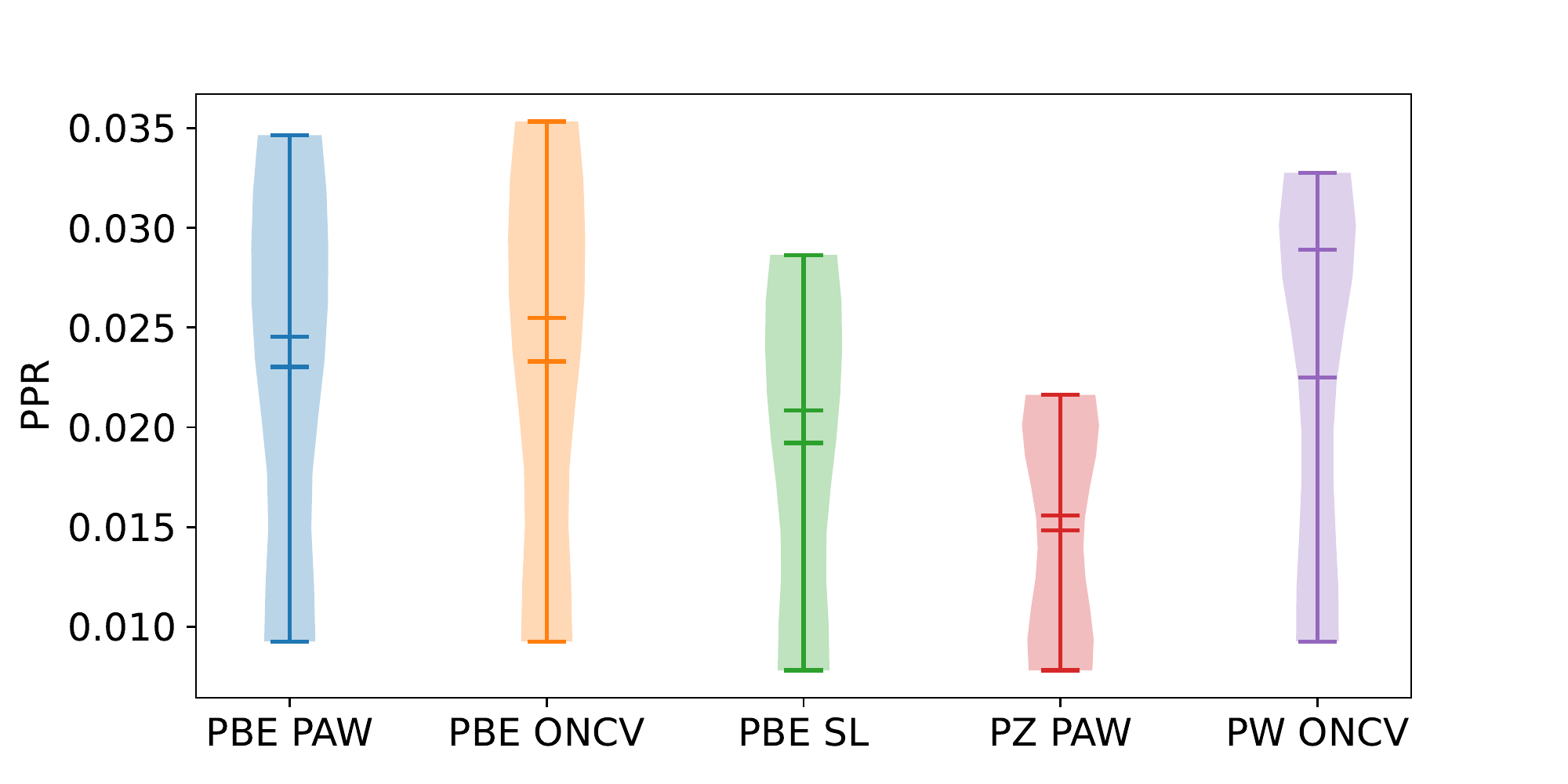}
         %\caption{}
         \label{fig:PR_antib}
     \end{subfigure}
     \caption{Violin plots showing (top) the participation ratio of all the WFs in all eigenenergies and (bottom) the participation ratio of the Sn(s,p) and S(p) states in the antibonding region (III) including the first four conduction band states. The two middle lines show the mean and median of the values.}
     \label{fig:PR}
\end{figure}

To quantify that change in the localization properties of the states, and of hybridization, the participation ratio of the MLWF is extracted. This is defined as \cite{Rehn2015, Fan2021},

\begin{equation}
\mathrm{PR} \left(n\right) = \frac{\left[\sum_{i=1}^N |\braket{i|n, \mathbf{k}=0}|^2\right]^2}{\sum_{i=1}^N |\braket{i|n, \mathbf{k}=0}|^4} 
\label{eq:PR}
\end{equation} 

where N is the total number of MLWFs included in the wannierization and n is the selected orbital state, while \textbf{k} = 0 denotes the home unit cell. The value of the participation ratio decreases with increasing localization, and it more properly reflects the number of orbitals over which hybridization of a state occurs. In analogy to Ref. \cite{Merkel2023}, we also define the `projected participation ratio', in the subspace W\textsubscript{s} of the antibonding states. For this, the states of eq. \ref{eq:PR} were renormallized, so as to be defined only in the subspace $i\in W_s$ of the 8 bands around the bandgap. The cut-off for this was taken so that no states below region III in fig. \ref{fig:PDOS} were taken. 

The results are shown in fig. \ref{fig:PR}. As the symmetry increases, the states become less hybridized overall (PR), or more localized. In the bottom panel (PPR), only the Sn(s,p) and the S(p) states, which participate in the lone pair model, are sampled. It is seen that the largest contribution to this trend in localization comes from these states, projected in the anti-bonding subspace. The RLP model is correctly recovered, which states that hybridization in the distorted structures is higher. As was mentioned in section \ref{sec:relaxation}, by using the PW ONCV PP, the Sn-S distances reduce significantly, and the hybridization between all neighboring atoms is wrongly emerging from the increase in coordination. In this case, the high PR and PPR ratios reflect this nonphysically higher coordination. The wannierization of this PP also did not converge, producing high Im/Re ratios and higher spread values. The slow convergence behavior of this PP, is another reflection of the high hybridization level occurring. The rest of the PPs, showed very good to excellent convergence behavior.  

\section{Born effective charges}

The dynamical atomic charges (Born Effective charges, Z\textsuperscript{*}) are considered next \cite{Dick1958, Ghosez1998}. The bulk periodic structure used here is not prone to depolarization effects, so it is easier to compare between different pseudopotential approximations. We calculated the Born effective charges of the Sn and S atoms using the Berry-phase theory of polarization \cite{Vanderbilt2000} in Quantum-Espresso \cite{Giannozzi2017, Umari2002}. An electric field of 0.036 V/$\AA$ was used, which is the same order of magnitude as the electronic gap in most cases ($\approx$0.01 V/$\AA$) \cite{Umari2002}.

\begin{figure*}
   \centering
     \includegraphics[width=.68\textwidth]{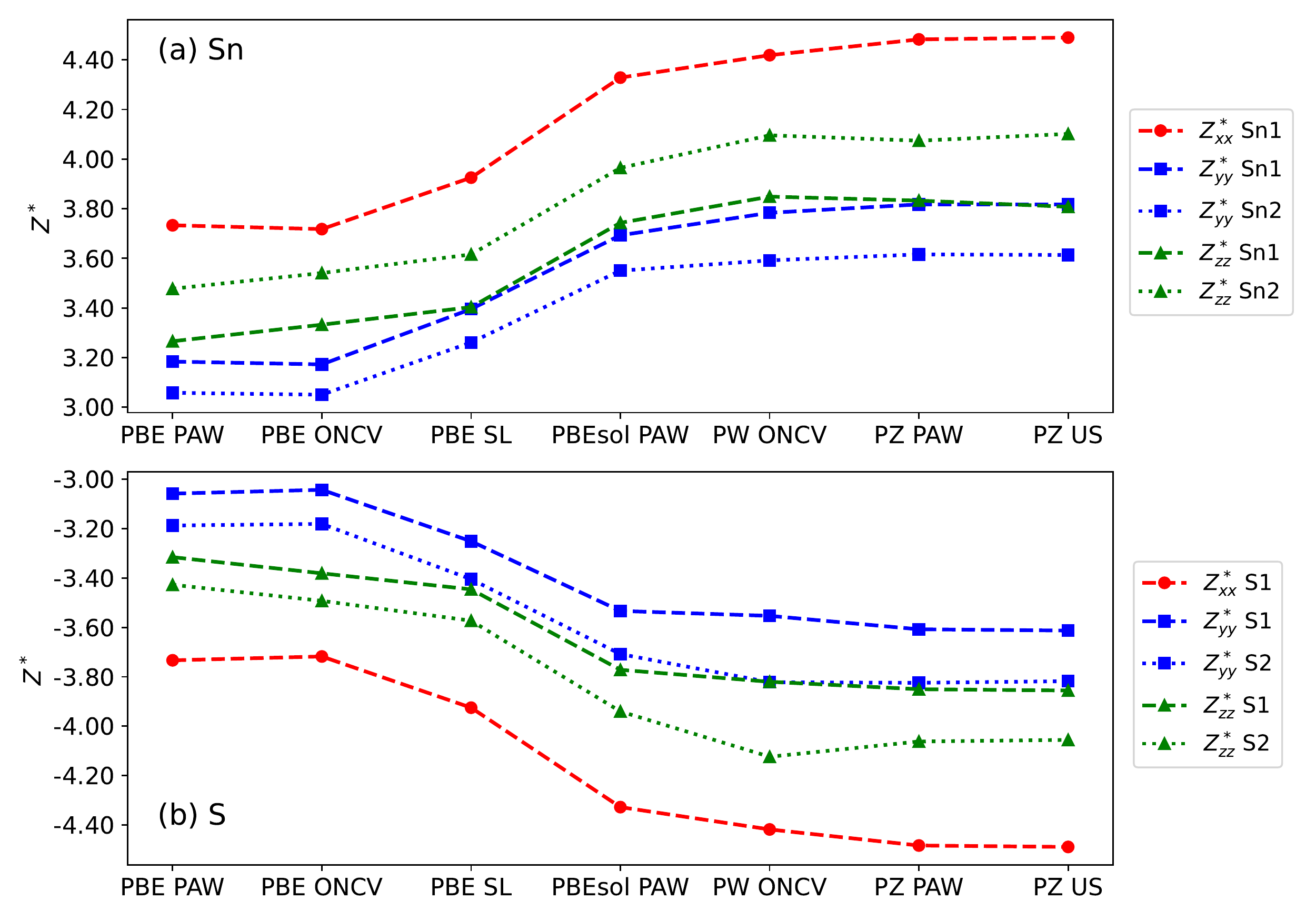}
     \caption{Born effective charges of the Sn and S atoms separately, for each PP. Sn1 and Sn2 refer to the bottom Sn atom of the top monolayer and the top Sn atom of the bottom monolayer respectively.  S2 is the in-plane neighbor of Sn1 and S1 of Sn2.}
         \label{fig:Born}
\end{figure*}  

\begin{figure*}
   \centering
     \includegraphics[width=.68\textwidth]{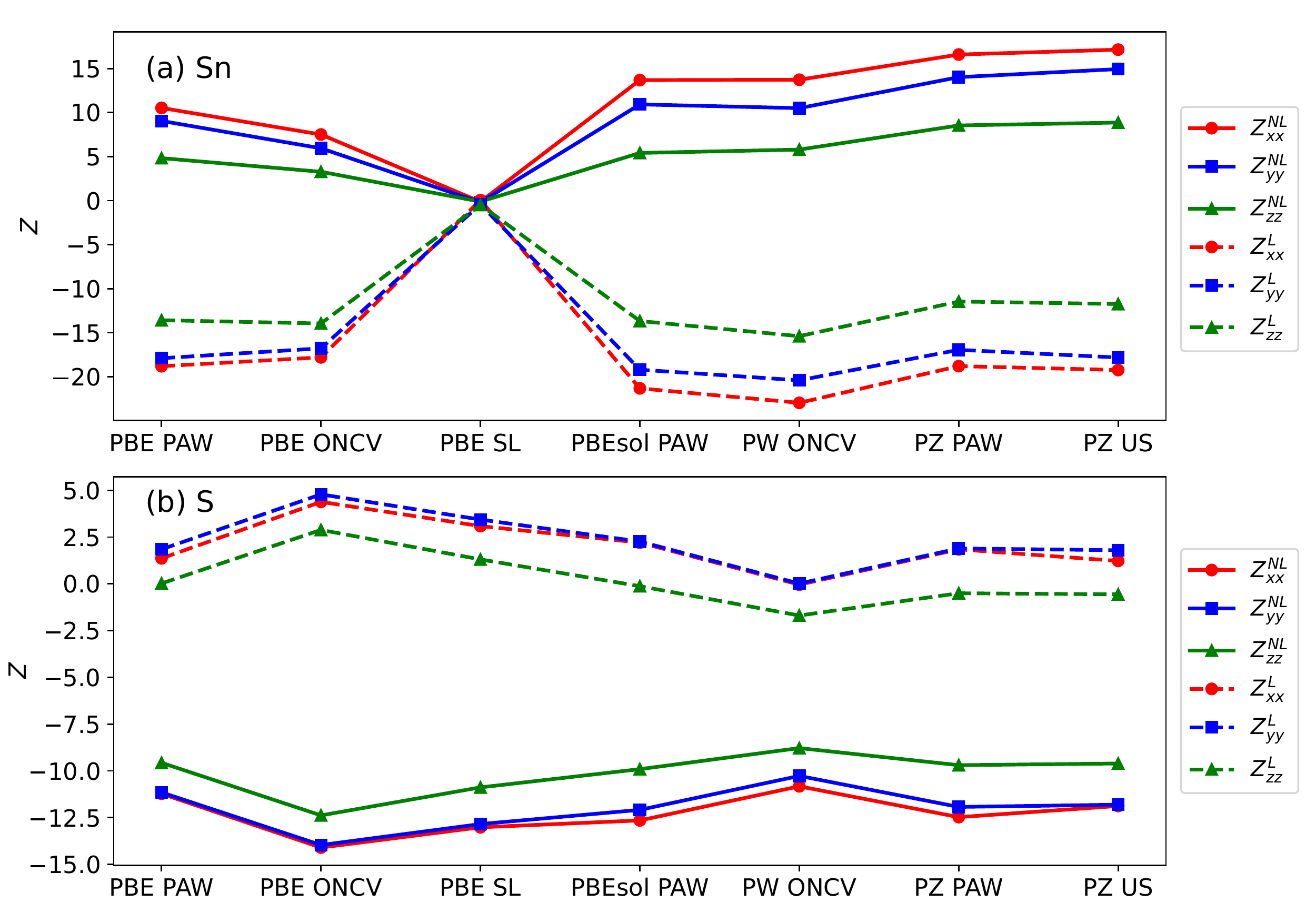}
     \caption{Components of the Born effective charges of the Sn and S atoms and separating the local and non-local contributions of the pseudopotential.}
         \label{fig:Born_nl}
\end{figure*}  

\begin{figure}
   \centering
     \includegraphics[width=.48\textwidth]{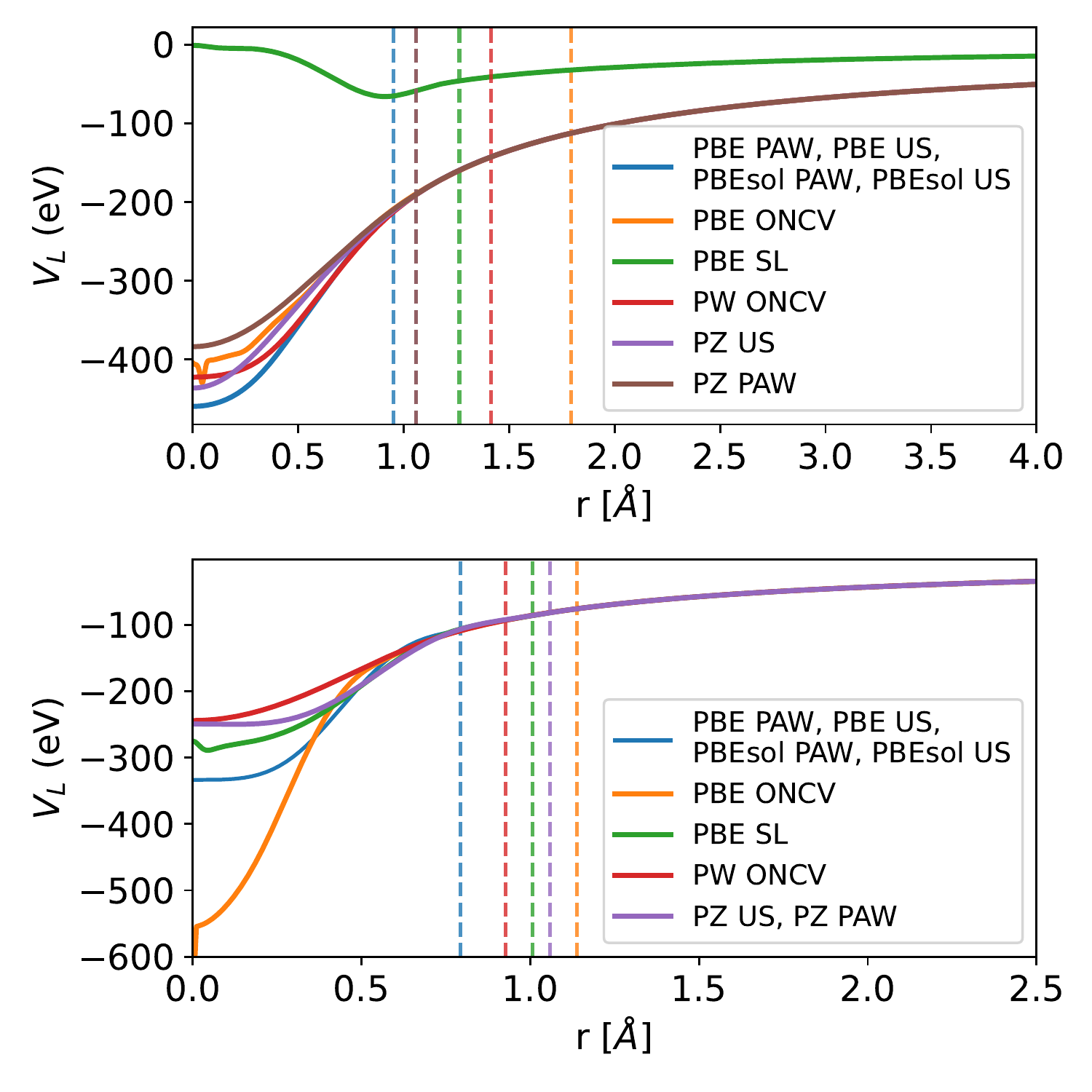}
     \caption{Local potential for the Sn (top) and S (bottom) atoms, as derived from the pseudopotential file.}
         \label{fig:Vpseudo}
\end{figure}  

The results are shown in fig. \ref{fig:Born}. We follow the same symmetry arguments as in the case of the WFs and only present `non-repeated' results. Our results for ONCV-PBE are close to those presented for bulk SnS in Ref. \cite{Dewandre2019} for the same PP. The same trend is observed in the description of the Z\textsuperscript{*} values, as in Figure \ref{fig:pressurePP} for the ratios of the lattice constants to a very good approximation. PBE PPs have lower values, which increase when passing to PBEsol and LDA functionals. The two latter have almost equivalent Z\textsuperscript{*} values.

The local (Z\textsuperscript{L}) and non-local (Z\textsuperscript{NL}) contributions for the atoms to the Z\textsuperscript{*} values were also extracted from the forces on the ions using equations \ref{eq:Z} and \ref{eq:forces}. This is shown in fig. \ref{fig:Born_nl}. In general, the magnitude of the contributions of both V\textsubscript{L} and V\textsubscript{NL} in the Sn atoms is lower for the PBE than the rest of the functionals, and higher for the S atoms. Especially in the case of SL, both local and non-local contributions for Sn go to zero.

To be able to explain the differences in the functional, the information on V\textsubscript{L} and V\textsubscript{H} around each atom, would suffice. This comes from the fact that the Z\textsuperscript{*} values are a result of the interplay of short-range and long range interactions. Then, equation \ref{eq:SR-LR} indicates that the local and Hartree contributions are enough to describe the long-range forces, if we are able to divide the two contributions.

The visualization of V\textsubscript{H} is an involved process, and the results can not be easily understood through the human eye, but would be more appropriate for machine-learning methods. On the contrary, some insight can quickly be gained directly from V\textsubscript{L} due to its rotational symmetry, and due to the fact that it is available before the self-consistent procedure. By looking at its two parts (short and long range) through the approximate division using the cut-off radius, we can understand the differences at least between the PBE functionals. This is shown in fig. \ref{fig:Vpseudo}, and in comparison to the LDA functionals, in order to use them as a second reference for our study.

Notably, we observe that the V\textsubscript{L} for the Sn atom with the SL PP goes to zero close to the core and is much shallower than the rest of the PPs. This explains the $Z_{x,y,z}^{L} \approx 0$ observed in fig. \ref{fig:Born_nl}.  For PAW and US using the PBE and PBEsol functionals, V\textsubscript{L} is equal. In these cases, it is the functional that dictates the origin of the differences in the Z\textsuperscript{*} values. The same is true for the two PZ PPs examined, but only for the case of the S atom. The last notable feature is in the case of PBE ONCV for the S atom. Here, the large dip of V\textsubscript{L} is translated into large Z\textsuperscript{L} values. 

Overall, we were able to trace back many of the origins of the Z\textsuperscript{*} values, which makes our analysis interesting for physically-motivated derivation of machine learning functionals and comparison with feature selection during machine learning pseudopotentials for dynamical lattice parameters and ferroelectric applications.

 \section*{Conclusions}
 
 We have successfully shown the dependence of the static and dynamic lattice properties to different types of pseudopotentials.
 
 The PPs could be categorized based on the \textit{a/b} lattice constant ratio they produced. The categorization also produced the same trend in the results for the density of states, when approximately dividing the energies in regions that belong to the bonding, intermediate and anti-bonding regions. This reflects the dependence of the density functionals in the revised lone pair model that dictates the stability of the layered structure to the lone pairs in between the layers. The dependence was further revealed by the participation ratio of the MLWFs, which point to higher hybridization for higher distortion, and also showed that nonphysical bonding produced by the density functional and PP combination had very  bad convergence during wannierization.

In terms of the dynamical lattice properties, the trend in the resulting Z\textsuperscript{*} values also followed the categorization that the PPs produced for the \textit{a/b} ratios. We traced back the properties of the local and non-local potential contributions to the Z\textsuperscript{*} values and also showed the connection of the former to V\textsubscript{L}.

One very interesting direction of this is the use of the results to train a model that learns dynamical lattice properties that can be measured experimentally and then build the functional by using experimental results \cite{Li2016, Brockherde2017}. The combination of short-range interactions of the local potential of the PP and the short-range Hartree interactions could then be used for this purpose, aiming at displacive ferroelectric materials.

 \section*{Methods}

 DFT calculations were performed via the \textsc{quantum espresso} package \cite{Giannozzi2017}. The pseudopotentials used were: 1) Ultra-soft, scalar-relativistic, Troullier-Martins \cite{Troullier1991} PBE PAW with non-linear core correction \cite{Louie1982} of Kresse-Joubert type \cite{Kresse1999} (PBE PAW), 2) Ultra-soft, scalar-relativistic PBE by Rappe, Rabe, Kaxiras and Joannopoulos (RRKJUS) \cite{Rappe1990, Rappe1991} with non-linear core correction, 3) Ultra-soft, scalar-relativistic PBEsol RRKJUS with non-linear core correction (PBEsol US) \cite{Perdew2008}, 4) Ultrasoft, scalar-relativistic PBEsol PAW with non-linear core correction (PBEsol PAW), 5) Semilocal, scalar-relativistic, Troullier-Martins PBE, 6) Scalar-relativistic, optimized norm-conserving Vanderbilt PBE pseudopotential (PBE ONCV), 7) LDA, scalar-relativistic, optimized norm-conserving Vanderbilt PW \cite{Perdew1992} (PW ONCV), 8) Ultra-soft LDA, scalar-relativistic PAW with non-linear core correction and self-interaction correction (PZ PAW) \cite{Perdew1981} and 9) Ultra-soft, scalar-relativistic, LDA of Troullier-Martins type (LDA US) with non-linear core correction \cite{DalCorso2014}.
 
 The parametrization for all pseudopotentials is as performed in the PSlibrary v1.0.0 by Dal Corso \cite{DalCorso2014}, except for the all ONCV type and the LDA US, which were parametrized by Hamman \cite{Hamann2013,Hamann2017}, while the PBE SL is taken from the FHI98PP package (PBE SL) \cite{Fuchs1999}. The forces acting on the atoms were relaxed to a value below 2.5 meV/$\mathrm{\AA}$, however, the results with the chosen e\textsubscript{T} did not change significantly when relaxing to a value up to three orders of magnitude lower for most PPs. For the case were Van der Walls corrections were included, this was done via the semi-empirical DFT-D3 method by Grimme \cite{Grimme2006}.

 The aiida software \cite{Vitale2020, Huber2020} was used for deriving the projectabilities of the states for the SCDM method. The convergence for the projected WFs (both with and without localization) was set to be achieved when both the gauge invariant and the gauge non-invariant parts of the total spread between successive iterations was less than 10\textsuperscript{-9} - 10\textsuperscript{-10} $\AA^2$, depending on the PP used. The imaginary/real ratios were mostly zero or below 10\textsuperscript{-5}, except for PZ PAW, where convergence was not achieved with these settings. A k-point of 11x11x11 was used in all cases. TBmodels was used for the extraction of the hopping integrals from the tight binding model of the wannier functions \cite{Gresch2017,Gresch2018}.

 \section{Acknowledgment}

Results presented in this work have been produced using the AUTH Compute Infrastructure and Resources and was supported by computational time granted from the Greek Research \& Technology Network (GRNET) in the `ARIS’ National HPC infrastructure under the project NOUS (pr012041).
      
\nocite{*}

\bibliography{apssamp}% Produces the bibliography via BibTeX.

\clearpage

\begin{widetext}

\section*{Supplementary Information}

\subsection{PDOS}

\begin{figure*}[b]
   \centering
     \begin{subfigure}{0.56\textwidth}
         \centering
         \includegraphics[width=\textwidth]{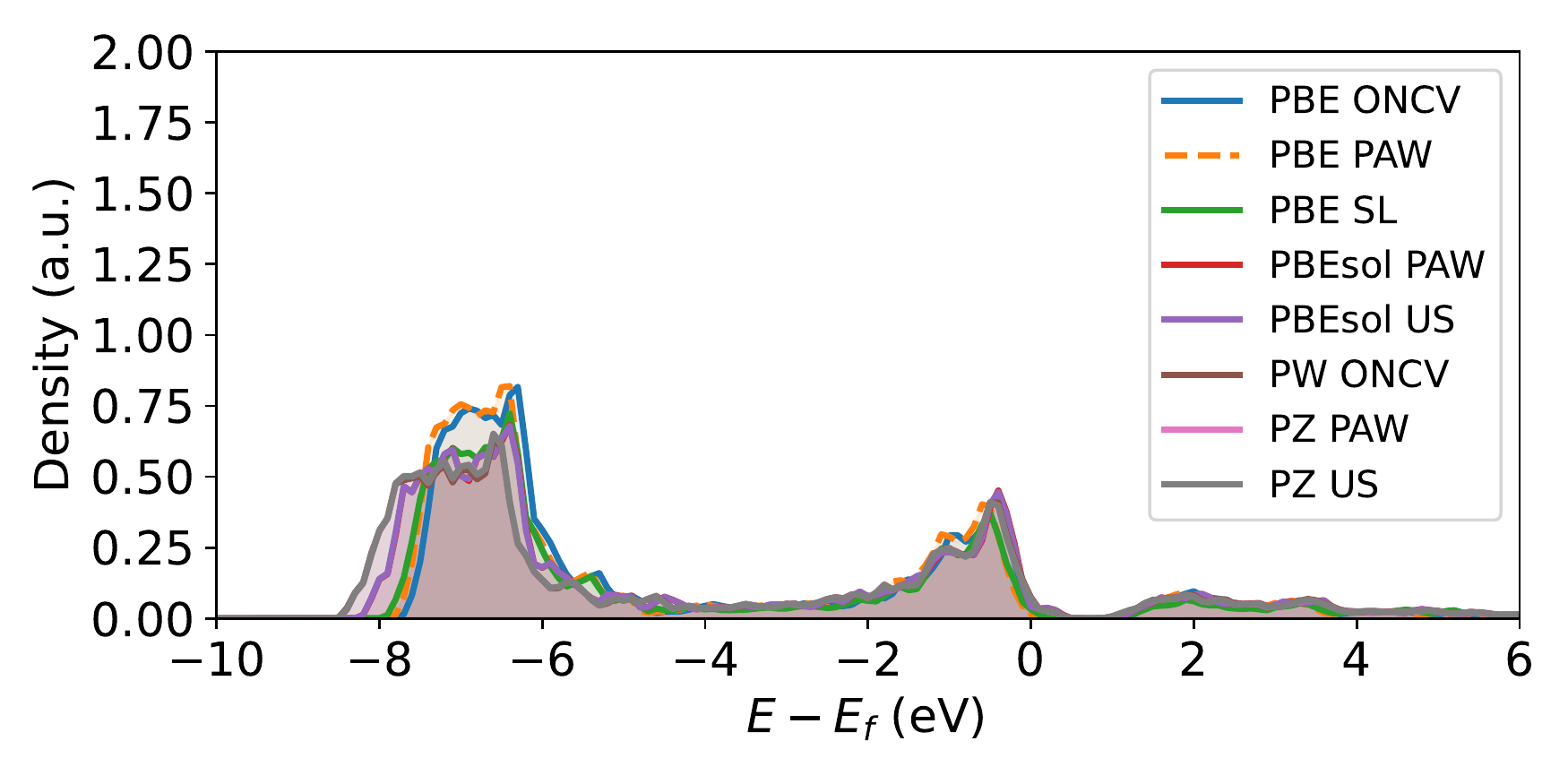}
         \caption{}
         \label{fig:Sn_s_all}
     \end{subfigure}
    \centering
    \hfill
     \begin{subfigure}{0.56\textwidth}
         \centering
         \includegraphics[width=\textwidth]{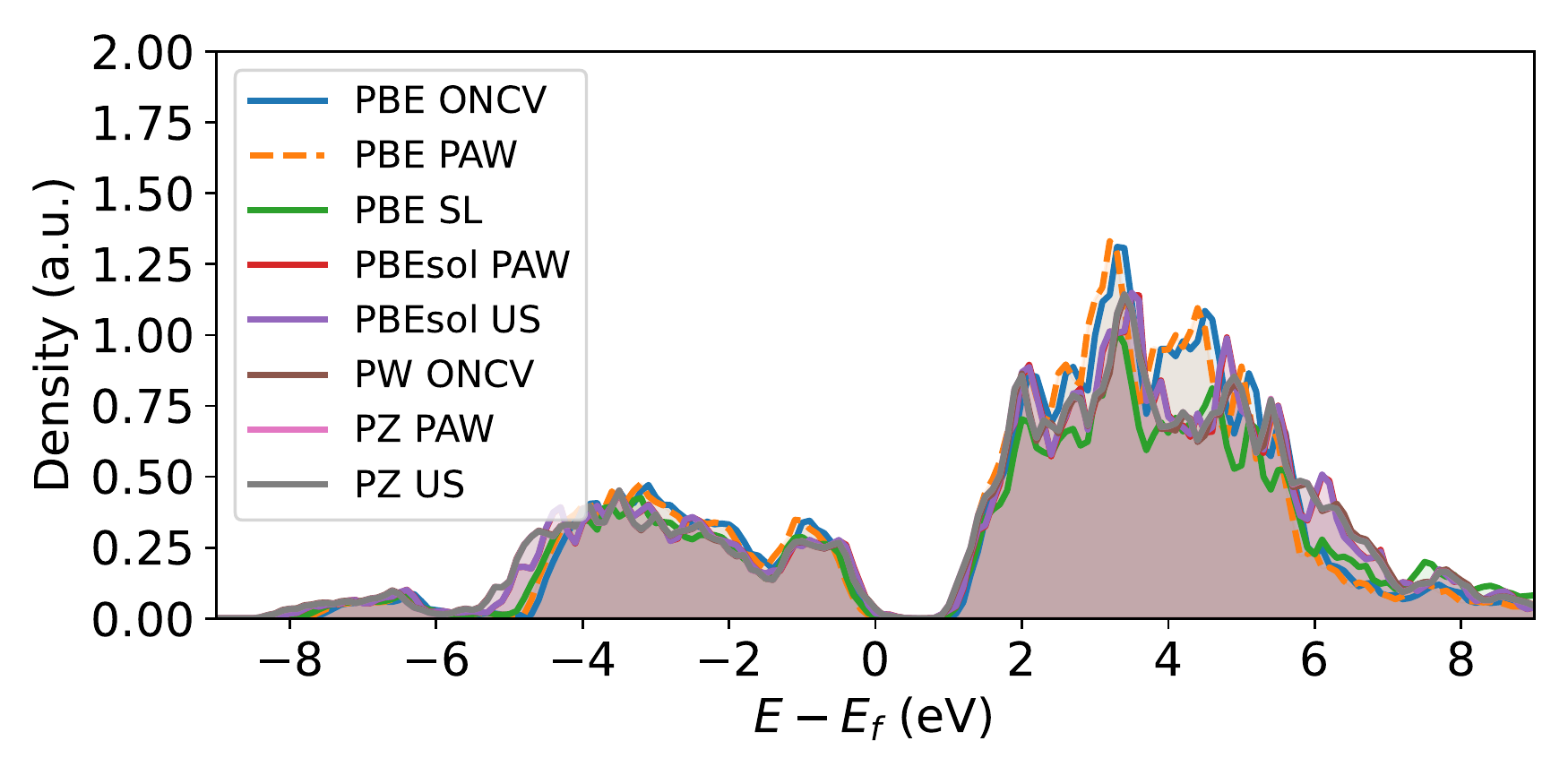}
         \caption{}
         \label{fig:Sn_p_all}
     \end{subfigure}
    \centering
    \hfill
     \begin{subfigure}{0.56\textwidth}
         \centering
         \includegraphics[width=\textwidth]{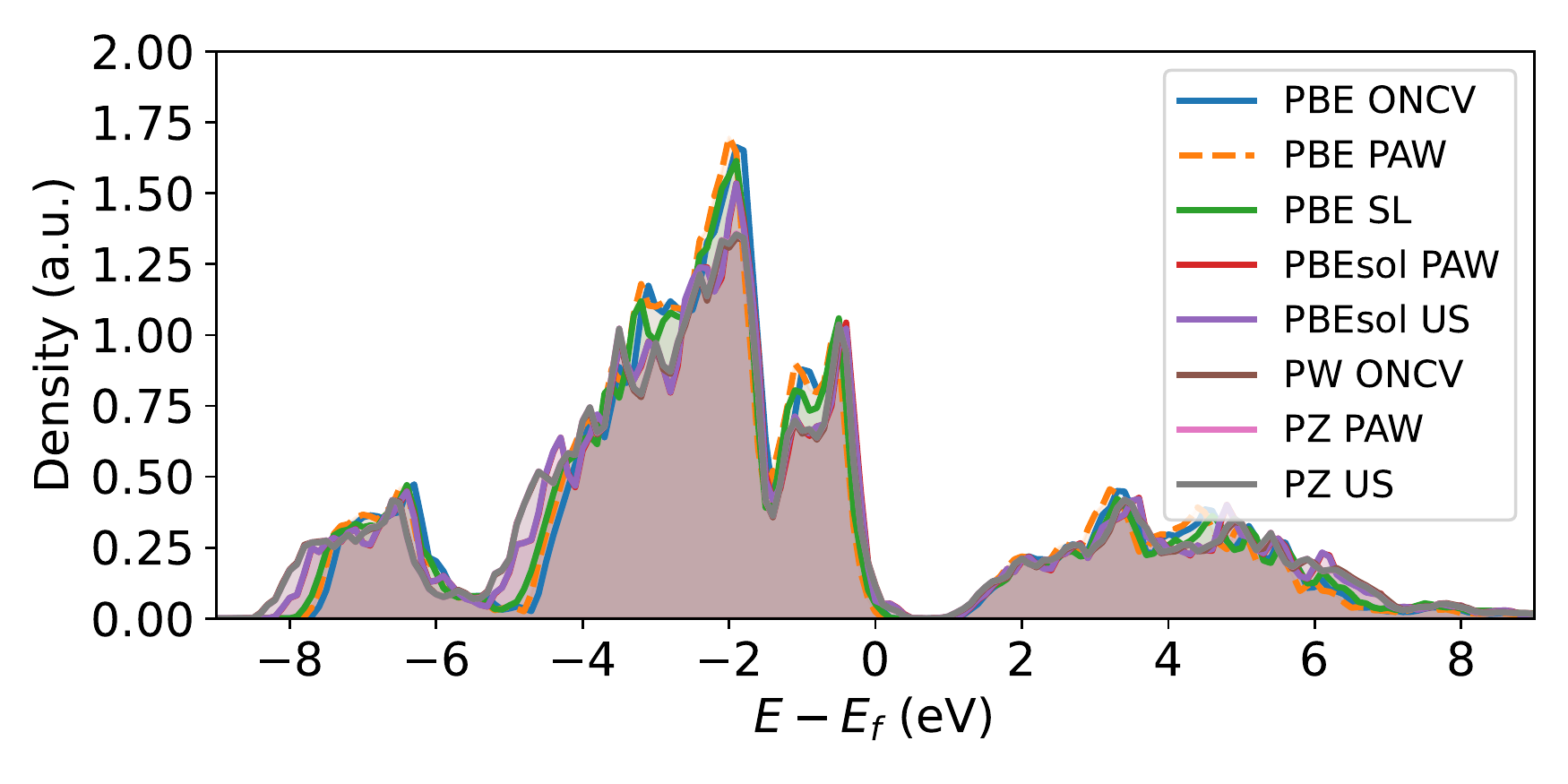}
         \caption{}
         \label{fig:S_p_all}
     \end{subfigure}
     \caption{Partial Density of states for the (a) Sn(s), (b) Sn(p) and (c) S(p) orbitals for the different pseudopotentials used in this work (see methods section in the main text for shorthand notation). Except for the SL, all pseudopotentials with PBE functional are almost overlapping, and similarly for the PBEsol and LDA approximations.}
     \label{fig:OPDOS_all}
\end{figure*}

\clearpage

\subsection{Wannierization}
\label{sec:wanniersup}

Figure \ref{fig:Projectabilities} shows the projectabilities of the states for SnS, as derived for PBE SL, using the aiida software. The complementary error function (CEF) is,

\begin{equation}
    f\left(e\right) = \frac{1}{2} \operatorname{erfc}\left(\frac{\epsilon - \mu}{\sigma}\right)
\end{equation}

The states for which the projectability is close to one are well-represented. Higher states (i.e. states with many radial nodes) are less easy to have a well-localized wannier representation. The wannierization with these parameters resulted in some localized states being outside the home unit cell and not related to the lone pair location. Therefore, these basis states were not used in this work.

\begin{figure}[h]
   \centering
     \includegraphics[width=.72\textwidth]{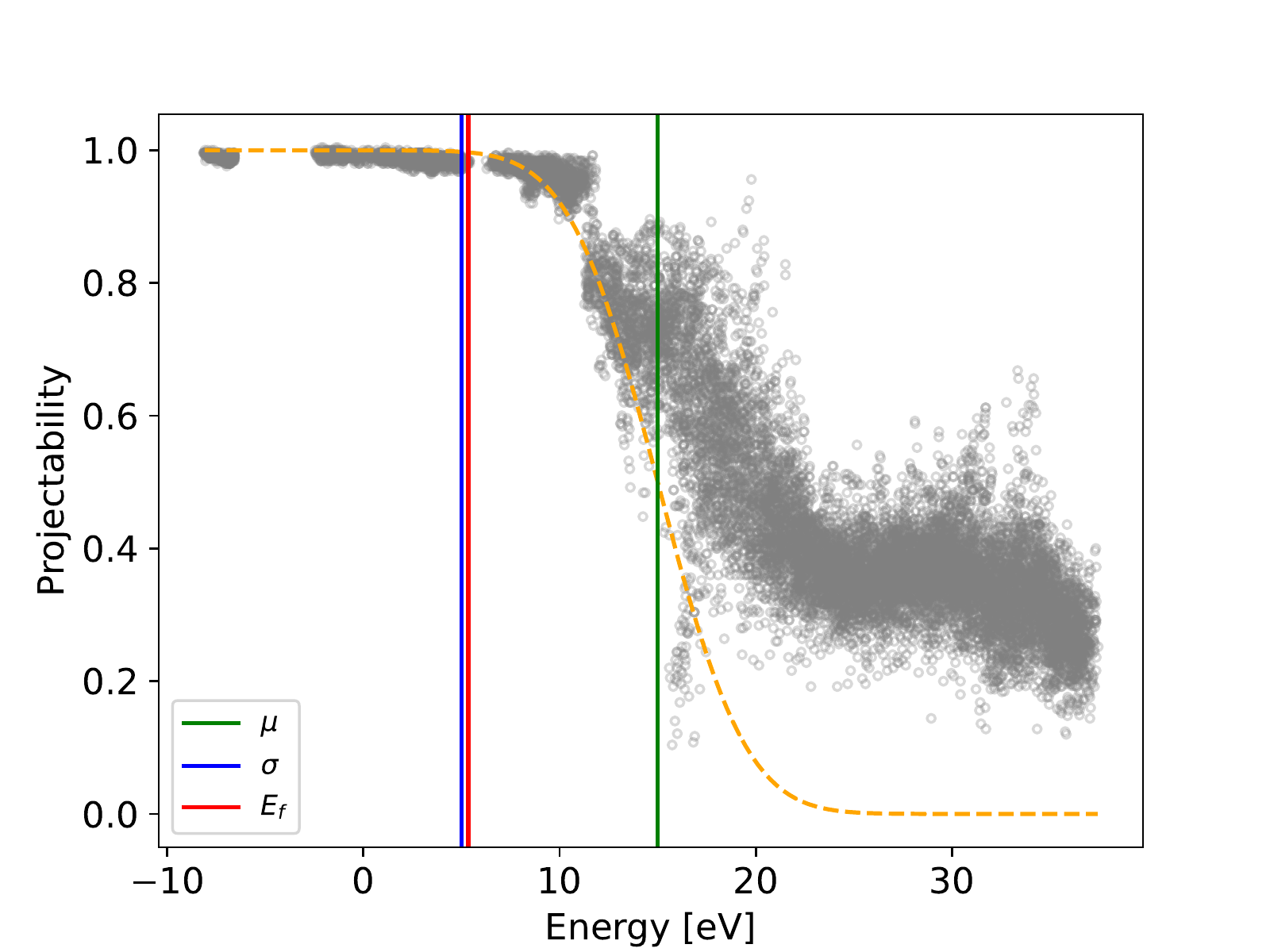}
     \caption{Projectabilities of states as a function of the energy (N\textsubscript{\textbf{k}}$\times$N\textsubscript{bands} points) for the PBE SL PP. The dashed line represents the CEF. $\mu$ and $\sigma$ correspond to one choice of optimum parameter that can be used for the CEF.}
         \label{fig:Projectabilities}
\end{figure}  

\clearpage

\end{widetext}

\end{document}